# Towards the Recapitulation of Ancient History in the Laboratory: Combining Synthetic Biology with Experimental Evolution

Betül Kaçar[1,2] and Eric Gaucher[2,3]

[1]NASA Astrobiology Institute, USA
[2]School of Biology, Georgia Institute of Technology, Atlanta, GA, 30322, USA
[3]School of Chemistry, Parker H. Petit Institute of Bioengineering and Biosciences,
Georgia Institute of Technology, Atlanta, GA, 30332, USA
betul@gatech.edu

## Abstract

One way to understand the role history plays on evolutionary trajectories is by giving ancient life a second opportunity to evolve. Our ability to empirically perform such an experiment, however, is limited by current experimental designs. Combining ancestral sequence reconstruction with synthetic biology allows us to resurrect the past within a modern context and has expanded our understanding of protein functionality within a historical context. Experimental evolution, on the other hand, provides us with the ability to study evolution in action, under controlled conditions in the laboratory. Here we describe a novel experimental setup that integrates two disparate fields - ancestral sequence reconstruction and experimental evolution. This allows us to rewind and replay the evolutionary history of ancient biomolecules in the laboratory. We anticipate that our combination will provide a deeper understanding of the underlying roles that contingency and determinism play in shaping evolutionary processes.

## Introduction

Living organisms are the product of their histories. Evolutionary biology is therefore an inherently historical science yet many details of this history are unobtainable: the fossil record is incomplete; ancestral genomic sequence information has been over-written via mutations; natural evolution occurs on long time scales; and the connections between genotype and phenotype are often intractable. Understanding these details is particularly difficult when one considers the potential role that chance plays in evolutionary outcomes. Along these lines, Stephen Jay Gould once remarked:

> [H]istory includes too much contingency, or shaping of present results by long chains of unpredictable antecedent states, rather than immediate determination by timeless laws of nature… (Gould 1994).

Gould's remark suggests that there are too many solutions for life to be repeatable. Such a suggestion implies that historical contingency is a fundamental determinant of evolutionary outcomes. Others, such as Simon Conway Morris, have argued that evolution is actually highly constrained, with many available pathways to only a relatively few destinations (Morris 2003). Advances in the field of experimental evolution and whole-genome sequencing now make it possible to empirically examine the role of historical contingency in evolution at both the organismal (Wichman et al. 2000; Counago et al. 2006; Blount et al. 2008; Pena et al. 2010; Meyer et al. 2012) and molecular levels (Weinreich et al. 2006; Poelwijk et al. 2007; Pennisi 2011; Salverda et al. 2011).

While various experimental evolution approaches have made much progress in dissecting the role of history in evolution by directly observing evolution in action, less is known about the direct relationship between genotypes (modern or ancient) and their effect on shaping an organism's evolutionary trajectory. Here we propose a novel synthesis of synthetic biology and experimental evolution that will further our understanding by combining molecular and systems evolution and provide an unprecedented means of addressing how contingency and deterministic forces interact to guide evolutionary trajectories.

## Rebuilding History and Creating Novelty with Synthetic Biology

Synthetic biologists assemble DNA to construct novel genes, metabolic pathways and even organisms (Benner and Sismour 2005; Endy 2005; Gibson et al. 2010). These manipulations provide us with a level of control that natural systems cannot provide, and this level of control minimizes unknown variables/parameters that effect particular systems. A powerful and increasingly useful synthetic biological approach is the computational reconstruction of ancient sequences of biomolecules using Ancestral Sequence Reconstruction (ASR), an approach sometimes referred to as paleogenetics. Initially proposed by Pauling and Zuckerkandl (Pauling and Zuckerkandl 1963), ASR merges history with natural selection (Stackhouse et al. 1990). ASR involves the alignment of DNA or protein sequences, followed by the construction of a phylogenetic tree that is then used to infer sequences of ancestral genes at interior nodes of a tree using likelihood and/or Bayesian statistics (Gaucher 2007). Recent advances in DNA synthesis now permit us to resurrect these ancient sequences in the laboratory and recombinantly express the ancient genes using modern organisms *in vivo* or reconstituted *in vitro* translation systems. Through a bottom-up approach we can engineer novel artificial systems that can be manipulated to better understand nature. The growing list of resurrected biomolecules now includes hormone receptors (Thornton et al. 2003), alcohol dehydrogenases (Thomson et al. 2005), elongation factors (Gaucher et al. 2008), thioredoxins (Perez-Jimenez et al. 2011), among others (Benner et al. 2007) and most recently, complex molecular machines (Finnigan et al. 2012).

As ASR follows a bottom-up approach and utilizes modern sequences to infer the past states of biomolecules, experimental evolution pursues a top-down approach that involves the real-time examination of the evolution of microbial model systems (Figure 1). (in the present context, top-down refers to complex cellular systems and/or to whole organisms). Experimental evolution has been used to address important questions in evolutionary biology (Elena and Lenski 2003). The experimental evolution approach is particularly powerful because of the high level of control it permits, the tractability of its microbial participants, and the capacity to create and maintain a viable frozen fossil record of the evolving populations that may then be used for highly detailed studies to address a variety of questions in evolutionary biology.

Here we introduce for the first time a novel system in which ASR is combined with experimental evolution, we term *paleo-experimental evolution*. In this approach, ASR is used to reconstruct an ancestral gene/protein. The synthetic ancestral gene is then used to precisely replace the endogenous form of the gene from a modern organism at the exact same chromosomal location. In some instances, we expect this replacement to cause the modern organism to be maladapted because the ancient gene/protein is not functionally equivalent to its (modern) descendent homolog. This synthetic recombinant organism is then experimentally evolved in the laboratory, and the subsequent adaptations are monitored using fitness measurements and whole-genome sequencing.

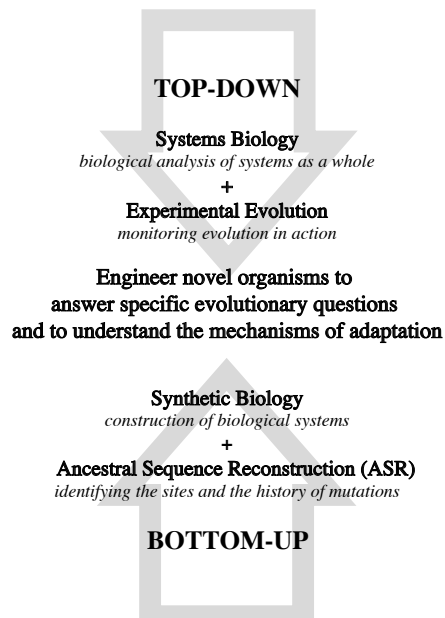

Figure 1: Artificial biology meets nature. In a novel paleo-experimental evolution system, descriptive evolutionary biology (top-down) meets applied, engineered synthetic biology (bottom-up) to further our understanding of evolutionary mechanisms

A paleo-experimental evolution setup also allows us to rewind and replay the molecular tape of life (or more precisely, one biomolecular component of life) to understand the role of chance and determinism in evolution, albeit in a laboratory setting. If evolutionary outcomes are deterministic, placing ancestral proteins within a modern context may result in the convergence of the ancient sequence towards the sequence of its modern counterpart. Alternatively, were historical contingency to be a major determinant of organismal evolution, there should be a number of available fitness peaks that may or may not be equally optimal and accessible via multiple trajectories. A major challenge, however, lies in our ability to develop a system that permits adaptation to occur along both deterministic and contingent paths if given equal *a priori* opportunity. Of course it is difficult to conceive of such an ideal system. However, we should be able to manage some aspects of such a system. For instance, if we choose to evolve an ancient enzyme that binds to only a single substrate and converts that substrate into a product subsequently used downstream in a metabolic pathway, then we are limited in the trajectories that the ancient enzyme can adapt (the enzyme can evolve or the substrate can change). On the other hand, if we choose to evolve an ancient enzyme that has numerous substrates and binds many ancillary protein partners, then we can expect such a system to evolve more contingently than the previous scenario because there is greater opportunity for compensatory co-evolution to overcome the low fitness of the ancestral protein when placed in a modern context. Again, the enzyme may evolve or the enzyme's substrate may change. Unique to this scenario, however, is the potential for interacting protein partners to accumulate mutations that restore interactions otherwise diminished by the ancestral protein.

### Ancient Hubs in Modern Times

A paleo-experimental evolution system that combines synthetic and evolutionary biology requires a deep understanding of the interactions of cellular components, biological networks, and gene regulation and expression. These components are shaped by the interplay between genotype and phenotype – the major determinants of natural selection.

What is fascinating in this complex picture is the harmonious *dialect,* a manner of language defined by intermolecular interactions within the context of the cell and that has the ability to respond and adapt to varying environments (Dennett 1995). Such a dialect can be a fine-tuned product of millions of years of evolutionary history both between and within the components of a cellular system. This very point challenges our ability to design new biological partners: fundamentally we are restricted by an organism's past.

Interchanging a modern protein in a cell with its homologous counterpart from another species can provide insight into the evolutionary paths and constraints that shape the evolution of homologous proteins. However, this intriguing experiment can fail to capture that the two homologs do not have a direct line of descent that *connects* them in progressive, linear time. Meaning, the evolutionary path that connects the two homologs requires that we travel *back* in time from one descendent to the common ancestor and then *forward* in time to the other descendent. As such, the homologs share a common ancestor but the descendents of that common ancestor traversed two separate (and possibly non-interchangeable) paths of adaptation and random fixation. This raises the possibility that the homologs are not 'functionally equivalent' (Figure 2A).

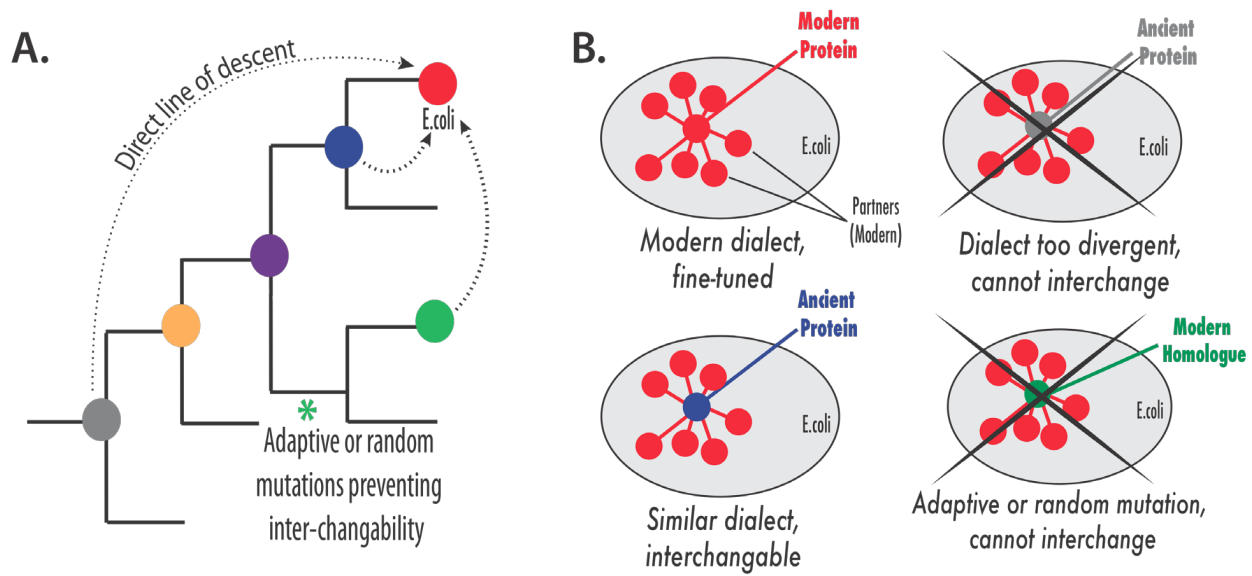

Figure 2. (A) Paleo-experimental evolution consists of resurrecting an ancient gene, removing the modern form of the gene from an extant organism, and then inserting the ancestral form into the extant organism. For instance, the ancient gene from the gray node on the phylogeny can be resurrected and then inserted into the *E. coli* genome (red node) at the precise chromosomal location that the extant gene was knocked out. This synthetic/engineered organism is then evolved in the laboratory. Our approach contrasts to other approaches that are only able to use modern genes from an organism to replace its ortholog in a different extant organism (say,

inserting the gene from the extant organism at the green node into the *E. coli*, red, organism). Such an approach can be limiting if adaptive or neutral mutations that prevent interoperability occurred along particular branches that connect the green and red nodes. (B) Protein interaction network containing modern and ancient hubs. Consider a particular hub protein that interacts with seven ancillary partners in *E. coli* (upper left). These interactions are fine-tuned over the course of evolution. Replacing the modern hub of the network with a recent ancestor of *E. coli* (blue) may permit the interaction network to still function, likely in a diminished capacity. Replacing the modern hub of the network with an ancient ancestor of *E. coli* (gray) may prevent the ancillary proteins from interacting with the hub altogether. Similarly, replacing the modern hub of the network with a divergent modern counterpart may prevent the interaction network from functioning despite that the same network exists in both modern organisms.

One manner in which homologs can become functionally nonequivalent is if a protein is part of a highly integrated molecular network in which the protein interacts with numerous ancillary partners. Sufficient co-adaptation or compensatory co-evolution amongst the protein and its ancillary partners along any phylogenetic lineage may prevent that particular protein from binding its necessary ancillary partners when interchanged in a different species (Figure 2B).

Replacing network partners with their ancestors would permit us to rewire a network within the historical context from which the mutational differences between the modern and ancestral proteins share a direct connection in evolutionary time. In a scenario where the hub (center) and nodes (terminal) of an interaction network have adapted to a particular lineage-specific dialect, replacing a component of such a network with its ancient counterpart may be analogous to resurrecting an old dialect that can be understood by it descendant speakers. As expected, the ability of the rest of the network to communicate (function) with this component from an ancient dialect would be limited by the manner in which the network changed between the ancestor and its modern form. The question of interest to us is whether the different components are capable of communicating or whether they will fail to communicate and thus be functionless – whoa is the Tower of Babel. We anticipate that ancestral components will in fact be able to communicate in the hub better than modern components from different species as long as the ancestor lies along the evolutionary path that directly connects the two modern proteins.

Despite our optimism, we suspect that the ancestral component will trigger a stress or strain on the modern network since the ancestral protein comes from an ancient dialect. If so, this creates an ideal scenario to watch the ancient component adapt within a modern network. Four possible scenarios may arise from such a system:

1) The ancient protein repeatedly adapts to the modern network in a manner identical or different to how its modern counterpart evolved (determinism).

2) The ancient protein adapts to the modern network in a manner different than how its descendent did (contingency).

3) The modern network adapts to the ancient protein in a manner identical to the ancient network - thus resurrecting the ancient network.

4) The modern network adapts to the ancient protein in a manner never evolved before in nature – thus creating an entirely new dialect.

The ability to differentiate these scenarios will determine the value of our paleo-experimental evolution system.

## An example paleo-experimental evolution system

Among the various proteins so far studied by paleogeneticists, Elongation Factor-Tu (EF) is an ideal candidate for use in paleo-experimental evolution. EF is a GTP-binding protein that functions to deliver aminoacylated-tRNAs to the A-site of the ribosome and is thus an essential component of ribosome-based protein biosynthesis (Czworkowski and Moore 1996). In addition to binding all ~47 different tRNAs (at least in *E. coli)*, EFs also bind to other classes of proteins such as chaperones, metabolic enzymes, structural proteins, and others (Figure 3). EFs are one of the most abundant proteins in bacteria. In addition to being a universal protein found in all known cellular life, deletion of EF is lethal (Schnell et al. 2003).

Previous studies using large protein datasets have calculated a correlation coefficient of 0.91 between environmental temperature of a host organism and the melting temperatures of a subset of a host's globular proteins (Gromiha et al. 1999). Among this subset of Towards the Recapitulation of Ancient History in the Laboratory: Combining Synthetic Biology with Experimental Evolution proteins, EFs are known to adapt to the environmental temperature of their host organisms - EFs from thermophilic microorganisms are thermostable whereas EFs from mesophilic organisms are mesostable; supporting the notion that proteins are marginally stable (Taverna and Goldstein 2002). This suggests that a strong selective constraint shapes the thermostability profile of EF proteins.

All these properties make EF-Tu an ideal protein for a system that combines experimental evolution with synthetic biology. The combination of EF's role in cellular networks and the strong constraints acting on EF's thermostability, creates an ideal situation to knockout endogenous EF from a modern organism and replace it with an ancestral form of the protein whereby the ancestral protein shares a direct evolutionary history with the modern form of the protein. We therefore set out to generate a strain of modern bacteria (*E. coli*) in which we replaced the endogenous EF with an ancestral EF at the precise genomic location of the modern gene – thus using the modern promoter to drive expression.

gene using DNA recombineering technology (Datsenko and Wanner 2000). *E. coli* is unique among most bacteria in that it contains two genomic copies of EF (*tufA* and *tufB*, that differ from one another by a single amino acid). We elected to insert the ancient EF at the *tufB* genomic location since this region of the chromosome is less populated with open reading frames of other genes compared to the *tufA* location. As such, we first knocked out *tufA* and measured this effect on growth (Figure 4). As a control for comparative purposes, we also knocked out *tufB* in a separate strain to measure its effect on growth (Figure 4). Next, we precisely swapped *tufB* for an ancestral EF gene in the *tufA* knockout strain.

Our ancestral EF represents an ancestral γ-proteobacteria that is estimated to be on the order of 500 million years old (Battistuzzi et al. 2004) and has 21 out of 394 amino acids differences with *E. coli*'s *tufB*. This marks the first time an ancient gene has been genomically integrated in place of its modern counterpart within a contemporary organism. We next measured the cellular doubling time of the synthetic recombinant organism hosting the ancestral gene. Figure 4 shows that when replaced with the modern EF gene, the ancient EF gene extended the doubling time by approximately two-fold.

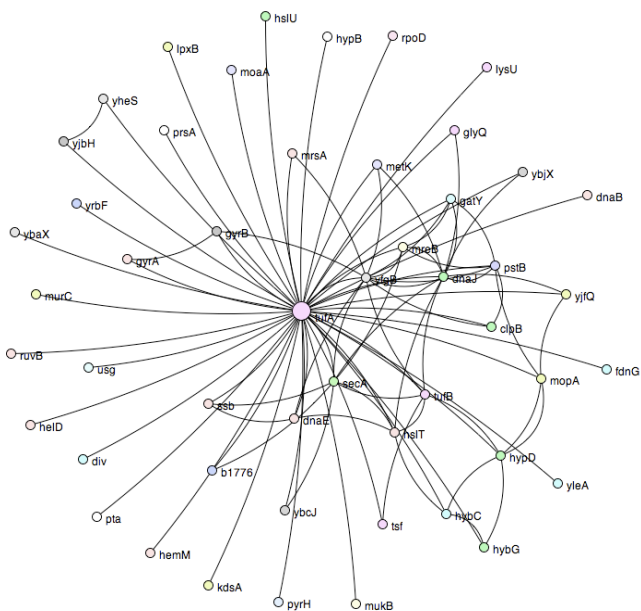

Figure 3: Bacterial EF-Tu (tufA node in center of hub) interacts with >100 cellular partners, including the ribosome, tRNAs, amino acids, GTP, EF-Ts (EF-Tu's nucleotide exchange factor) and more. This graph shows the >50 protein binding partners to EF-Tu that have been experimentally validated (binding to nucleic acids not shown). Network dataset rendered using Bacteriome.org.

To fulfill our paleo-experimental evolution objective in the laboratory, we have replaced the modern endogenous EF-Tu gene with a resurrected form of the

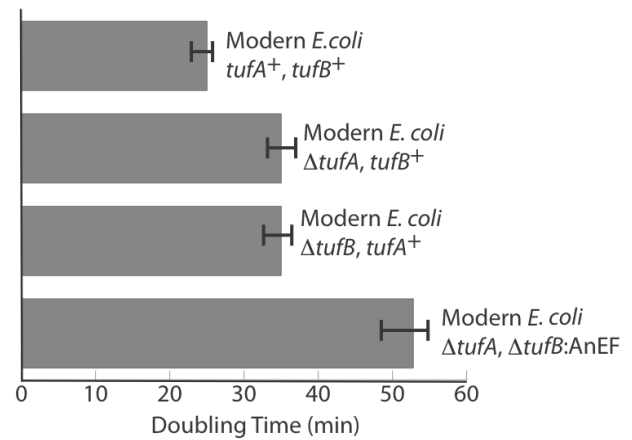

Figure 4: Precise replacement of a modern bacterial EF-Tu gene with its ~500 million year old ancestor extends the bacterial doubling time by two-fold. Two genes, *tufA* and *tufB*, (varying by just one amino acid) code for EF-Tus in *E. coli*. Precise replacement of endogenous EF-Tu requires both chromosomal *tufA* and *tufB* to be disrupted *(Schnell et al. 2003)*. Deletions of *tufA* or *tufB* in the *E. coli* B strain have similar effects (~ 34 minutes) when deleted individually. The ancient EF (AnEF) has 21 (out of 392) amino acid differences with the modern EF-Tu protein. Measurements are performed in LB media at 37°C in triplicate. Modern *E. coli* B strain REL606 was obtained courtesy of R. E. Lenski (Michigan State University).

## Historical contingency and the unpredictability of life

A paleo-experimental evolution system in the laboratory permits us to travel back in time to some approximation. By exploiting paleogenetics, we effectively go back in time through the history of a single component of life, capture that component, and transport it with us back to the present. In its most abstract manner, we have rewound a section of the tape of life and are giving it another opportunity to 'evolve' (albeit in a modern context). This approach therefore allows us to experimentally carryout Gould's thought experiment on "replaying the tape of life" at the molecular level. We anticipate that our novel system will enable us to address long-standing questions in evolutionary and molecular biology:

- Does an organism's history constrain its future?
- Does evolution always lead to a single and defined point or are there multiple solutions?
- How does a gene network adapt (as a whole or individual nodes)?
- Are compensatory mutations predictable?
- How do gene networks affect the evolutionary trajectory of a whole genome?
- How does selection act at the level of gene regulation vs. protein behavior?
- What is the impact of epistasis in shaping adaptive landscapes?
- Do universal biological laws govern evolution?

In addition to the points above, we anticipate that our system will enable us to address issues regarding the predictability of evolution. Along these lines, three important factors necessary to predict evolutionary outcomes are evolutionary dynamics, evolutionary rates and understanding the constraints acting on an evolving system. Changing the connectivity of a protein interaction network by swapping the network's hub with its various evolutionary ancestors provides us with an opportunity to control some of these factors and may lead to predictability at some level. For instance, we can control the amount of stress or strain on our synthetic recombinant organism by controlling the ancestral hub we introduce into the system. Older, more ancient EFs are expected to be a greater burden when placed in a modern organism compared to an ancient EF resurrected from a node closer on a tree to the modern organism. In a system where evolutionary stressors can be controlled, how much of evolution will follow random paths? If the evolutionary trajectories are dependent on evolutionary starting points (different ancestral states), and if we can control the factors of an evolving system; will life follow an unpredictable path?

## Conclusion

In this article, we introduce and describe a novel experimental setup that we term paleo-experimental evolution. This setup weds synthetic biology with experimental evolution. The goal of this combination is to identify the historical stops along the evolutionary tracks that gave rise to modern genotypes and to explore the accessible peaks in evolutionary history, thus helping us determine the role of chance vs. necessity in evolution. Despite the unnatural properties of our laboratory system, we anticipate that our unique system will advance our ability to understand both evolutionary mechanisms and how genotype is connected to phenotype even when phenotype arises in a synthetic system.

It should be noted that our system is not limited to ancient genes. *De novo* genes can be engineered and placed in organisms as well and the evolutionary patterns that arise from their adaptation can be tested *in vivo*. Further, synthetic genes can be evolved in additional genomic backgrounds (e.g., a thermophilic and a mesophilic species) for a deeper understanding of the role that a genome's *history* has in shaping a synthetic gene's evolutionary trajectory when placed in a modern organism.

We anticipate that our ability to combine the two disparate fields of synthetic biology and experimental evolution will enhance our understanding of the constraints that shape biological evolution. If we are able to demonstrate that aspects of evolution are predictable regardless of whether this is due to strong selective constraints or due to historical events, this insight will be valuable in our ultimate attempts to generate artificial life and our ability to maintain (and when necessary, constrain) this life form.

## Acknowledgements

This work was funded by National Aeronautics and Space Agency Astrobiology: Exobiology and Evolutionary Biology grant NNX08AO12G to E.A.G and by the NASA Astrobiology Institute through a NASA Postdoctoral Fellowship to B.K.